\documentclass[epj]{webofc}
\usepackage[varg]{txfonts}   
\usepackage{color}
\usepackage{bbold}
\usepackage{capt-of}
\usepackage{pstricks}
\usepackage{bm} 
\usepackage{slashed}
\usepackage{booktabs}
\usepackage{tikz}
\usepackage{pgfplots}
\pgfplotsset{/pgf/number format/use comma,compat=newest}

\wocname{EPJ Web of Conferences}
\woctitle{CONF12}
\long\def\comment#1{ }

\def\x{{\boldsymbol x}}

\newcommand{\avg}[1]{\langle #1 \rangle}

\newcommand{\be}{\begin{equation}}
\newcommand{\ee}{\end{equation}}
\newcommand{\beq}{\begin{eqnarray}}
\newcommand{\eeq}{\end{eqnarray}}

\newcommand*\diff{\mathop{}\!\mathrm{d}}

\providecommand*{\eu}{\ensuremath{\mathrm{e}}}

\newcommand{\id}{\mathbb{1}}

\begin{document}
\selectlanguage{english}
\title{Parity doubling of nucleons, Delta and Omega baryons across the deconfinement phase transition}
%
%

\author{Gert Aarts\inst{1} \and
        Chris Allton\inst{1} \and
        Davide De Boni\inst{1}\fnsep\thanks{\email{d.de-boni.840671@swansea.ac.uk}} \and
        Simon Hands\inst{1} \and
        Chrisanthi Praki\inst{1} \and
        Benjamin J\"ager\inst{1,2}\fnsep\thanks{2 = current address} \and
        Jon-Ivar Skullerud\inst{3}
}

\institute{Department of Physics, College of Science, Swansea University, Swansea SA2 8PP, United Kingdom 
\and
           ETH Z\"urich, Institute for Theoretical Physics, Wolfgang-Pauli-Str. 27, 8093 Z\"urich, Switzerland 
\and
           Department of Mathematical Physics, National University of Ireland Maynooth, Maynooth, Country Kildare, Ireland
}

\abstract{%
In this work we analyse positive- and negative-parity channels for the nucleon (spin $1/2$ octet), $\Delta$ and $\Omega$ baryons (spin $3/2$ decuplet) using lattice QCD.
In Nature, at zero temperature, chiral symmetry is spontaneously broken, causing positive- and negative-parity ground states to have different masses.
However, chiral symmetry is expected to be restored (for massless quarks) around the crossover temperature, implying that the two opposite parity channels should become degenerate.
Here we study what happens in a temperature range wich includes both the hadronic and the quark gluon plasma (QGP) phase.
By analysing the correlation and spectral functions via exponential fits and the Maximum Entropy Method respectively,
we have found parity doubling for the nucleon and $\Delta$ baryon channels in the QGP phase.
For the $\Omega$ baryon we see a clear signal of parity doubling at the crossover 
temperature, which is however not complete, due to the nonzero strange quark mass. 
Moreover, in-medium effects in the hadronic phase are evident for all three baryons, in particular for the negative-parity ground states.
This might have implications for the hadron resonance gas model. 
In this work we used the FASTSUM anisotropic $N_f = 2 + 1$ ensembles.
}

\maketitle

\newpage 

\section{Introduction}
\label{intro}

In Nature, at zero temperature, a considerable mass difference between the negative-parity ground state of the baryons and the positive-parity one
is understood from chiral symmetry breaking.
In the case of the nucleon and $\Delta$ baryon, this mass difference is far too big to be explained by the small explicit breaking of chiral symmetry due to the light $u$ and $d$ quarks.
In fact it is well-known that the mass difference between the opposite-parity ground states is mainly a consequence of the spontaneous breaking of chiral symmetry.
Since in the case of massless quarks chiral symmetry is expected to be restored above the deconfinement temperature, one would expect to see
parity doubling in the QGP phase.\\
On the other hand, chiral symmetry restoration is not fully realised for the $\Omega$ baryon because of the relatively large mass of the strange quark.
Therefore it would be very interesting to investigate what happens to the opposite parity channels of this particle at high temperatures.
While there are many works on chiral symmetry at finite temperature in the mesonic sector (see e.g. $\!$\cite{Rapp}), surprisingly only a few quenched analyses
are available in the baryonic sector ~\cite{Kogut,Datta,Pushkina}. 
Our aim here is to analyse parity doubling in the unquenched baryonic sector, in particular for the nucleon, $\Delta$ and $\Omega$ baryons.
We study both correlators and spectral functions below and above the crossover temperature $T_c\,$.
Our previous analyses for the nucleon sector can be found in \cite{Aarts:2015mma,Aarts:2015xua} and, more recently, \cite{DeBoni:lat2016} for the $\Delta$ baryon.


\section{Baryonic correlators and spectral functions}
\label{sec-1}

In general a baryonic correlator is written as (see for instance ~\cite{Montvay:1994cy,Gattringer:2010zz})
\be\label{correlator}
C(x)=\avg{O^\alpha(x)\,\overline{O}^\alpha(0)}\:,
\ee
with an implicit sum over the spin index $\alpha\,$. The simplest annihilation operators for the nucleon, $\Delta$ and $\Omega$ baryons are respectively
\be
O_N^{\alpha}(x) =\epsilon_{abc}\,u_a^{\alpha}(x)\left(d^{^T}_b(x)\,C\,\gamma_5\,
u_c(x)\right)\:,
\ee
and ~\cite{Leinweber:2004it,Edwards:2004sx}
\be
O_\Delta^{\alpha}(x) =
\epsilon_{abc} \left[ 
2\, u_a^{\alpha}(x)\left(d^{^T}_b(x)\,C\,\gamma_i\, u_c(x)\right) +
d_a^{\alpha}(x)\left(u^{^T}_b(x)\,C\,\gamma_i\, u_c(x)\right) \right]\:,
\ee
\be
O_\Omega^{\alpha}(x) =
\epsilon_{abc} \left[ 
s_a^{\alpha}(x)\left(s^{^T}_b(x)\,C\,\gamma_i\, s_c(x)\right) \right]\:,
\ee
where the Lorentz index $i$ is not summed and $C$ corresponds to the charge conjugation matrix.
We then project to a definite parity state by taking into account the interpolator $O_{N^{\pm}}=P_{\pm}\,O_N$ (analogously for the $\Delta$ and $\Omega$ baryons) in (\ref{correlator}), where
\be
P_{\pm} = \frac{1}{2} \left( \id \pm \gamma_4 \right)
\ee
projects to positive or negative parity.
We consider solely zero three-momentum correlators 
\be
C_{\pm}(\tau) = \int\!\!\diff{\x}\,C_{\pm}(\tau,\x)\:.
\ee
Each correlator contains both parity channels since $C_{-}(\tau)=-C_{+}(1/T-\tau)\,$. This means that the positive-parity channel propagates forwards in time,
whereas the negative-parity one propagates backwards in time.
For massless quarks one can prove ~\cite{Gattringer:2010zz,wip} that a chiral rotation on the quark fields gives $C_{\pm}(\tau)=-C_{\mp}(\tau)\,$, implying that the two parity channels are degenerate.\\
Using the Maximum Entropy Method (MEM)~\cite{Asakawa:2000tr}, we reconstruct the baryonic spectral functions $\rho(\omega)\,$, 
which are related to the baryonic correlators through the spectral relation \cite{wip}
\be
C_{\pm}(\tau)=
\int_{-\infty}^{+\infty}\frac{\diff{\omega}}{2\pi}\,\rho_{\pm}(\omega)\,\frac{\eu^{-\omega\tau}}{1+\eu^{-\omega/T}}\:.
\ee
It is straightforward to prove that $\rho_{+}(-\omega)=-\rho_{-}(\omega)\,$, therefore positive (negative) frequencies of $\rho_+$ correspond to the positive (negative) parity channel.
Moreover $\rho_{\pm}(\omega)$ are even functions in case of parity doubling.
The MEM procedure demands that the spectral function be positive. One can prove that $\rho_+(\omega)\geq 0 \:\:\forall \omega\in\mathbb{R}\,$ ~\cite{wip}.


\section{Lattice setup}
\label{sec-2}

\begin{table}[t!]
\centering
\caption{Simulation parameters used in this work. The available statistics
for each ensemble is $N_{\rm cfg} \times N_{\rm src}$. The sources were chosen randomly in the four-dimensional lattice. The value $N_{\rm cfg}=139.5$ means that there are $139$ configurations with $16$ sources and $1$ with $8\,$.}
\begin{tabular}{cccccccc}
\hline
$N_s$ & $N_\tau$  & $T$\,[MeV] & $T/T_c$  & $N_{\rm src}$  & $N_{\rm cfg}$\\
\hline
  24 & 128 & 44	& 0.24  &  16 & 139.5\\
  24 & 40 & 141 & 0.76  &  4  & 501  \\
  24 & 36 & 156 & 0.84  &  4  & 501  \\
  24 & 32 & 176 & 0.95  &  2  & 1000 \\ 
  24 & 28 & 201 & 1.09  &  2  & 1001 \\ 
  24 & 24 & 235 & 1.27  &  2  & 1001 \\  
  24 & 20 & 281 & 1.52  &  2  & 1000 \\ 
  24 & 16 & 352 & 1.90  &  2  & 1001 \\ 
\hline
\end{tabular}
\label{tab:lat}
\end{table}

The configurations used here are created by the FASTSUM
collaboration~\cite{Aarts:2014cda,Aarts:2014nba,Amato:2013naa}, with 2+1
flavours of non-perturbatively-improved Wilson fermions. The configurations and the
correlation functions have been generated using the CHROMA software
package~\cite{Edwards:2004sx}, via the SSE
optimizations when possible \cite{McClendon}.
Tab.~\ref{tab:lat} shows the simulation parameters based on the setup of the Hadron Spectrum Collaboration~\cite{Edwards:2008ja}.
The masses of the $u$ and $d$ quarks produce an unphysical pion with a mass of $384(4)\,\mbox{MeV}$~\cite{Lin}. 
The strange quark has been tuned to its physical value, therefore we expect the mass of the $\Omega$ baryon to be close
to the physical one.
In order to better reconstruct the spectral function from the correlator, we used an anisotropic lattice
with $a_s/a_\tau = 3.5$ and $a_s=0.1227(8)\,\mbox{fm}\,$. This allows us to have a sufficiently large
number of points in the Euclidean time direction even at high temperatures.
From the calculation of the renormalized Polyakov loop one extracts the crossover temperature $T_c=183$ MeV,
which is higher than in Nature, due to the large pion mass.\\
Concerning the baryonic correlators, gaussian smearing~\cite{Gusken:1989ad} has been employed to increase the overlap
with the ground state. In order to have a positive spectral weight, we apply the
smearing on both source and sink, i.e. 
\be
\psi' = \frac{1}{A} \left(\id + \kappa \, H \right)^{n} \psi\:, 
\ee
where $A$ is an appropriate normalization and $H$ is the spatial hopping part of the Dirac operator.
We tuned the parameters to the values $n=60$ and $\kappa = 4.2\,$, for maximising
the length of the plateau for the effective mass of the ground state in the $N_s^3\times N_t=24^3 \times 128$ lattice.
The hopping term contains APE smeared links~\cite{Albanese:1987ds} using $\alpha =
1.33$ and one iteration. The smearing procedure is only used in the spatial directions and
applied equally to all temperatures and ensembles.


\section{Results for $N$, $\Delta$ and $\Omega$ baryons}
\label{sec-3}

\begin{figure}[t!]
\centering
\includegraphics[width=7cm,clip]{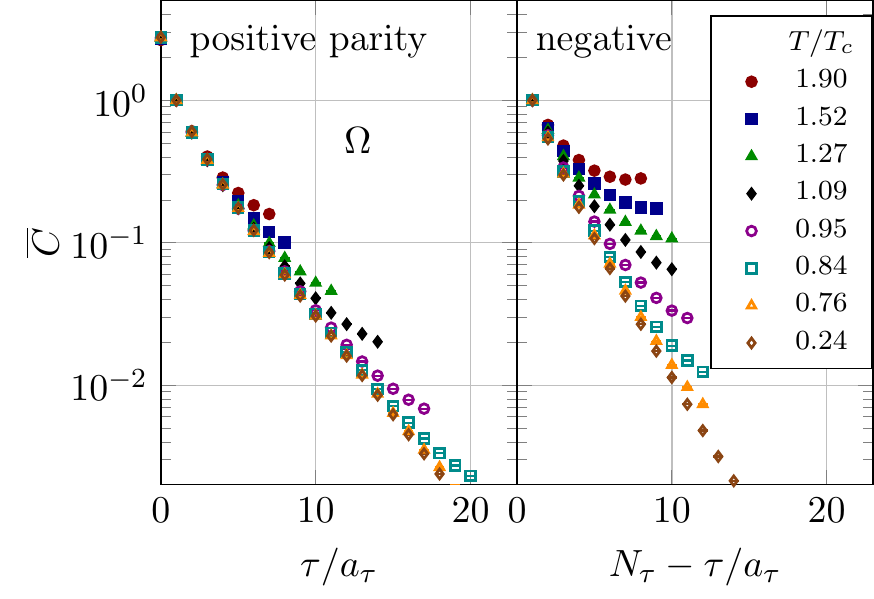}
\includegraphics[width=5.95cm,clip]{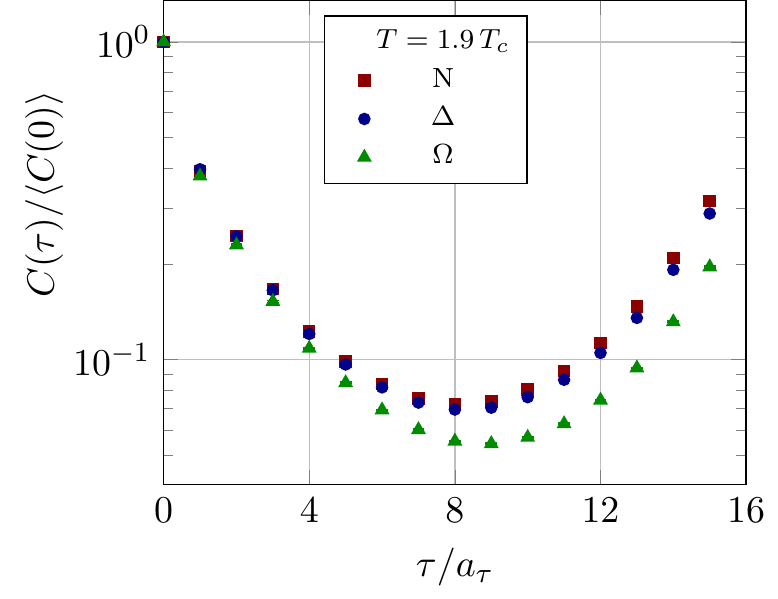}
\caption{On the left: $\Omega$-baryon correlators for the positive and negative parity channels at different temperatures.
On the right: Correlators of the nucleon, $\Delta$ and $\Omega$ baryons at the highest temperature for our lattice.}
\label{fig-1}       
\end{figure}

\begin{figure}[t!]
\centering
\includegraphics[width=7cm,clip]{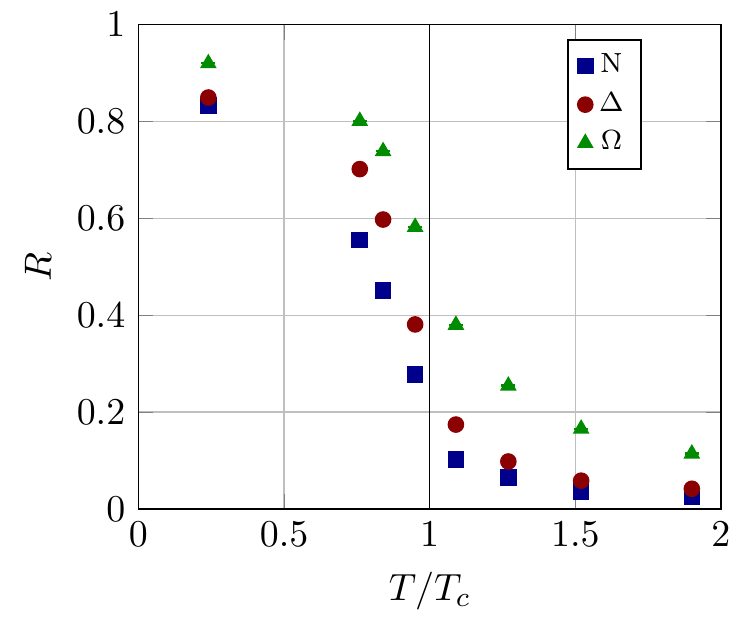}
\includegraphics[width=7cm,height=5.6cm,clip]{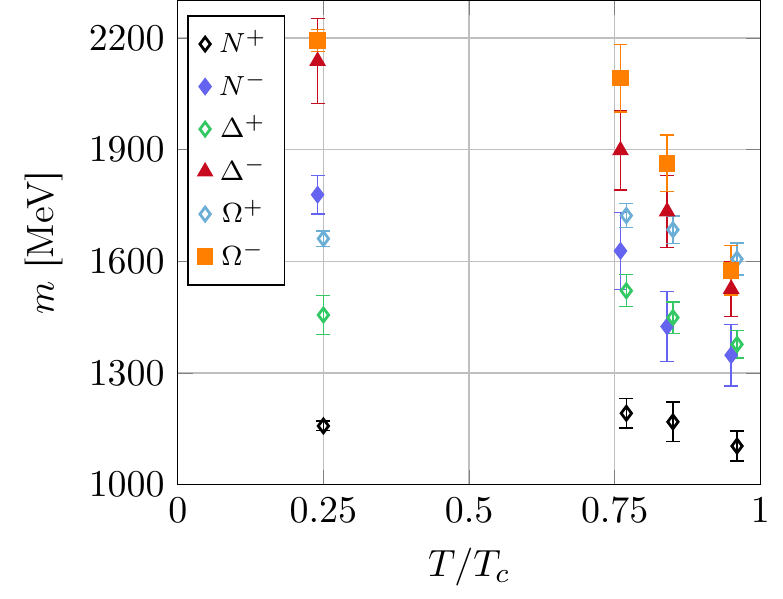}
\caption{On the left: Temperature dependence of the $R$ factors of the nucleon, $\Delta$ and $\Omega$ baryons. On the right: Ground state masses obtained using exponential fits to the nucleon, $\Delta$- and $\Omega$-baryon correlators at temperatures below $T_c\,$. The data for the positive-parity ground states are slightly shifted to the right in order to better compare with the other values.}
\label{fig-2}       
\end{figure}

The correlator of the $\Omega$ baryon is shown on the left panel of Fig.~\ref{fig-1},
in which the positive- and negative-parity channels are plotted separately.  
The correlators for the nucleon and $\Delta$ baryon are shown in~\cite{DeBoni:lat2016},
and their temperature behaviour is very similar to the one of the $\Omega$ baryon correlator. 
The correlators have been normalised to the first Euclidean time $\tau=a_\tau$
($\,\tau=N_\tau a_\tau -a_\tau\,$) for the positive-(negative-) parity partner, i.e.  (we write $C=C_+$ for ease of notation)
\be
 \overline{C}_{+}(\tau) = \frac{C\left( \tau \right)}{\langle C(a_\tau) \rangle} \quad
  \mbox{and} \quad
  \overline{C}_{-}(\tau) = \frac{C\left( N_\tau a_\tau- \tau \right)}{\langle C(N_\tau a_\tau - a_\tau)
  \rangle}.
\ee
This normalisation allows us to better compare the data at different temperatures. The right panel of Fig.~\ref{fig-1}
shows approximately symmetric correlators for both the nucleon and $\Delta$ baryon at the highest temperature we consider.
This means that the two parity channels are degenerate. Moreover, the $N$ and $\Delta$ correlators are almost identical,
suggesting that they represent quasi-free $u$ and $d$ quarks, and do not distinguish the different spin dependence in the two channels.\\  
The left panel of Fig.~\ref{fig-2} shows the summed ratios of the three baryons, defined as
\be
\label{eq:ratioR}
R \equiv
\frac{\sum_{n=1}^{N_\tau/
2-1}R(\tau_n)/\sigma^2(\tau_n)}{\sum_{n=1}^{N_\tau/2-1}1/\sigma^2(\tau_n)}\:,
\ee
where
\be
R(\tau)\equiv \frac{C(\tau)-C(N_\tau a_\tau-\tau)}{C(\tau)+C(N_\tau a_\tau-\tau)}\:.
\ee
\begin{table}[t!]
\centering
\begin{tabular}{cccccc}
$T/T_c$ & 0 [PDG] & 0.24 & 0.76 & 0.84 & 0.95 \\
\midrule
$m_N^+$ [MeV] & 939 & 1158(13) & 1192(39) & 1169(53) & 1104(40) \\
$m_N^-$ [MeV] & 1535(10)  & 1779(52) & 1628(104) & 1425(94) & 1348(83) \\
\midrule
$m_\Delta^+$ [MeV] & 1232(2) & 1456(53) & 1521(43) & 1449(42) & 1377(37) \\
$m_\Delta^-$ [MeV] & 1710(40) & 2138(114) & 1898(106) & 1734(97) & 1526(74) \\
\midrule
$m_\Omega^+$ [MeV] & 1672.4(0.3) & 1661(21) & 1723(32) & 1685(37) & 1606(43) \\
$m_\Omega^-$ [MeV] & 2250? 2380? 2470? & 2193(30) & 2092(91) & 1863(76) & 1576(66) \\
\midrule
$\delta_N$ & 0.241(1) & 0.212(15) & 0.155(35) & 0.099(40) & 0.100(35) \\
$\delta_\Delta$ & 0.162(14) & 0.190(31) & 0.110(31) & 0.089(31) & 0.051(28) \\
$\delta_\Omega$ & 0.147? 0.175? 0.192? & 0.138(9) & 0.097(23) & 0.050(23) & -0.009(25) \\
\end{tabular}
\caption{Ground state masses obtained using exponential fits to the nucleon, $\Delta$ and $\Omega$ baryons 
correlators for temperatures below $T_c\,$. The masses of the positive and negative parity ground states
include an estimate for statistical and systematic uncertainties.The ratios $\delta_N\,$, $\delta_\Delta$ and $\delta_\Omega$ are defined as $\delta=(m_--m_+)/(m_-+m_+)\,$.
Note that $\delta_\Omega$ is not accessible because $m_\Omega^-$ is still unknown.}
\label{tab}
\end{table}
\begin{figure}[t!]
\centering
\includegraphics[width=7cm,clip]{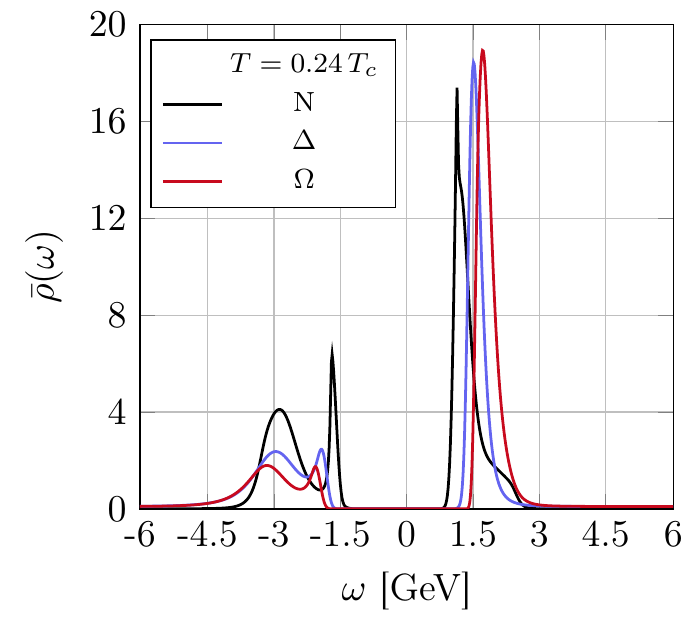}
\includegraphics[width=7cm,height=6.2cm,clip]{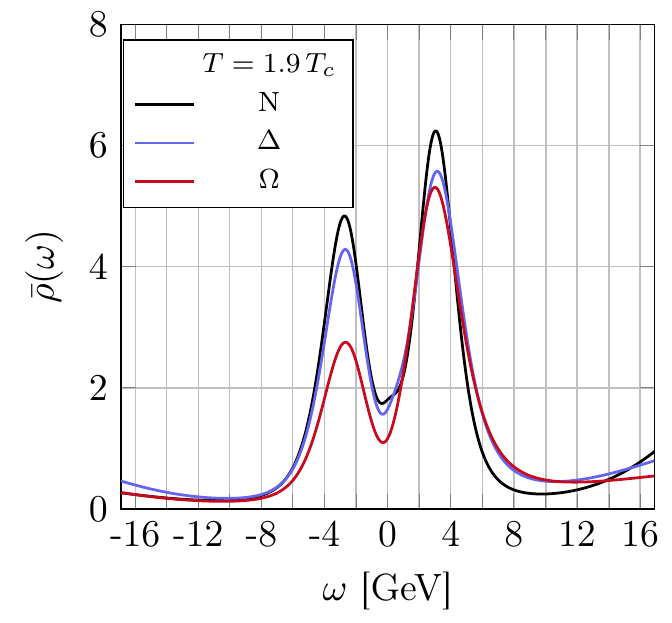}
\includegraphics[width=7cm,clip]{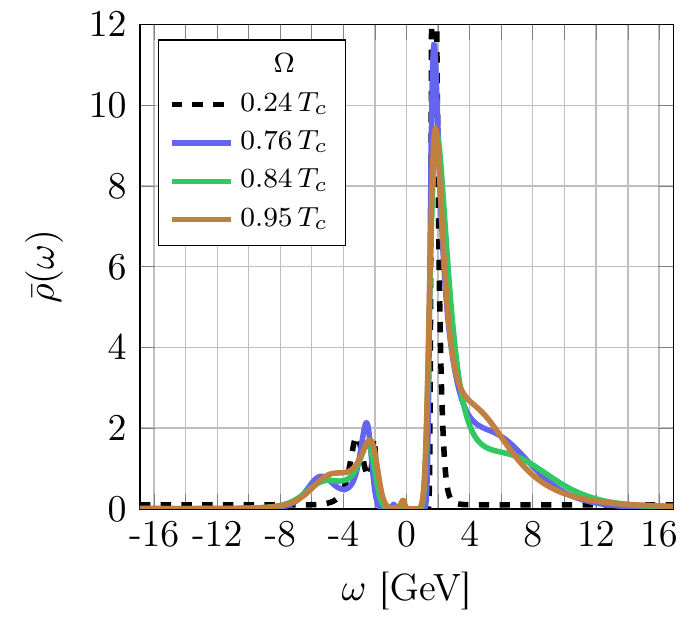}
\includegraphics[width=7cm,height=6.3cm,clip]{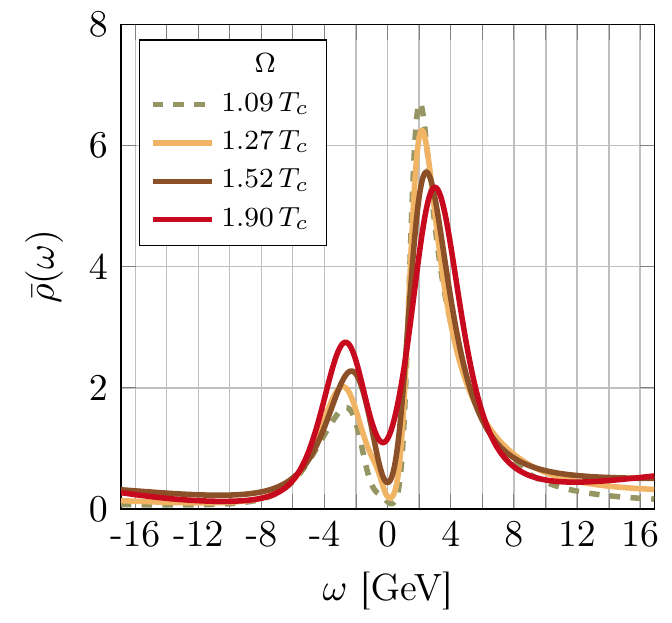}
\caption{Top: On the left (right), comparison between the normalised spectral functions obtained using MEM at the lowest (highest) temperature on our lattice for the three baryons.
Bottom: On the left (right), normalised spectral functions below (above) $T_c$ for the $\Omega$ baryon.
The dimensionless object $\bar\rho(\omega)$ is obtained from the normalised correlator $C(\tau)/(a_\tau\,\langle C(\tau=0)\rangle)\,$.}
\label{fig-3}       
\end{figure}
$\!$By definition the $R$ factor lies between $0$ and $1$, and $R=0$ corresponds to a symmetric correlator. We use the statistical 
uncertainties as weights in eq.(\ref{eq:ratioR}).
On the left of Fig.~\ref{fig-2} we see a clear signal of parity doubling around the crossover temperature $T_c$ for all the three baryons,
with possibly a slightly delayed effect for the $\Omega$ particle.
The $R$ factor is very close to zero at $T=1.9\,T_c$ for the nucleon and $\Delta$ baryon, indicating that their correlation functions
become almost symmetric in the QGP phase (as already shown on the right of Fig.~\ref{fig-1}).
On the other hand, the $R$ factor for the $\Omega$ baryon remains finite at our highest temperature. This is expected
from what is shown on the right plot of Fig.~\ref{fig-1}, in which the $\Omega$ correlator is still asymmetric.
This indicates that we do not have a complete parity doubling for the $\Omega$ baryon at these temperatures, due to the finite strange quark mass of approxiamately $100$ MeV.\\
In Tab.~\ref{tab} and in the right panel of Fig.~\ref{fig-2} we show the ground state masses in the confined phase extracted from 
a simple exponential fit of the correlation function, that is
\be
C_+(\tau)= A_{+}\,\eu^{-m_{+}\tau}+A_{-}\,\eu^{-m_{-}(N_\tau a_\tau-\tau)}\:. 
\label{eq:ansatz}
\ee
In order to estimate the systematic uncertainties of the
four fit parameters, we have considered various Euclidean time intervals. 
To further suppress excited states, we have excluded very small times.
The so-called Extended Frequentist Method~\cite{Yao:2006px,Durr} has been used for carrying out the statistical analysis.
This method considers all possible variations and weights
the final results according to the obtained $p$-value, which measures how extreme an outcome is. Further information on this
method can be found in~\cite{Yao:2006px,Durr}.\\
The lattice spacing was set by using the zero-temperature mass of the positive-parity ground state of the $\Omega$ baryon~\cite{Lin},
therefore, by construction, the value we found at $T=0.24\,T_c$ has to be in agreement with the value of $1672.4(0.3)$ MeV found in Nature~\cite{Agashe:2014kda}.
The ground state mass of the negative-parity channel is still unknown in the PDG and there are three possible candidates.
The value we obtained in Tab.~\ref{tab} at $T=0.24\,T_c$ seems to favour the candidate with the lowest mass.
However, a systematic analysis (continuum extrapolation and physical $u$ and $d$ quarks) is necessary to make a prediction.\\
One can see that in-medium effects are more important in the negative-parity channel for all three baryons, since the mass of the negative-parity
ground state decreases considerably when temperature is increased, whereas the mass of the positive-parity partner is almost unaffected
by temperature.
The spectral function of the $\Omega$ baryon at different temperatures is plotted in the lower panels of Fig.~\ref{fig-3}. A similar plot for the other two baryons can be
found in~\cite{DeBoni:lat2016}. The positive-parity channel corresponds to $\omega>0\,$, whereas $\omega<0$ refers to the negative-parity channel.
The spectral functions of the $\Omega$ baryon are not even functions of $\omega\,$, either below or above $T_c\,$, indicating that the opposite parity channels are not degenerate.
But one can still see a signal of parity doubling above $T_c\,$, since the spectral function becomes more symmetric with respect to the origin when
temperature increases. In order to have a clear plot of many spectral functions in the same figure, error bars are not displayed.
However, they do not modify what have been said above about the results.\\
At the top of Fig.~\ref{fig-3} we show a comparison between the spectral functions of the three baryons for the lowest and highest temperatures on our lattice.
Below $T_c$ all the spectral functions are very asymmetric, whereas above $T_c$ the spectral functions of the nucleon and $\Delta$ baryon are almost symmetric.
Moreover, as a consequence of the almost identical  nucleon and $\Delta$-baryon correlator at $T=1.9\,T_c\,$, the corresponding spectral functions are nearly identical.


\section{Conclusions}

By studying the temperature dependence of the correlators, spectral functions and $R$ factor, we clearly observe a signal of
parity doubling of the ground state across the crossover temperature for the nucleon and $\Delta$ baryon, with 
parity doubling realised almost completely at the highest temperature on our lattice. In the $\Omega$-baryon case
the opposite parity ground states remain still distinct, but we still observe a tendency towards parity doubling. 
For the nucleon and $\Delta$ particle, the asymmetry in the parity partners at zero temperature is mainly due to spontaneous breaking of chiral symmetry, 
hence the observed parity doubling can be understood from the restoration of chiral symmetry, which is expected to occur at high temperature.
We note that there is still a small explicit breaking of chiral symmetry since we are using
massive $u$ and $d$ quarks in the Wilson formulation.
On the other hand, the explicit breaking is not negligible in the case of the $\Omega$ baryon, which contains
$s$ quarks with physical mass of the order of $T_c\,$. Therefore the fact that parity doubling is not fully realised for this particle can be
understood from this explicit breaking of chiral symmetry due to the massive $s$ quark.


\section{Acknowledgements}

This work used the DiRAC Blue Gene Q Shared Petaflop system at the
University of Edinburgh, operated by the Edinburgh Parallel Computing
Centre on behalf of the STFC DiRAC HPC Facility (www.dirac.ac.uk). This
equipment was funded by BIS National E-infrastructure capital grant
ST/K000411/1, STFC capital grant ST/H008845/1, and STFC DiRAC Operations
grants ST/K005804/1 and ST/K005790/1. DiRAC is part of the National
E-Infrastructure. We acknowledge also STFC grant ST/L000369/1.
We are grateful to the Royal Society, the Wolfson Foundation and the Leverhulme Trust for support.



\newpage

\begin{thebibliography}{99}


\bibitem{Rapp}
R.~Rapp and J.~Wambach, \textbf{Adv. Nucl. Phys.} 25 (2000) 1,

\bibitem{Kogut}
C.~E.~DeTar and J.~B.~Kogut, \textbf{Phys. Rev. Lett.} 59 (1987) 399; \textbf{Phys. Rev. D} 36 (1987) 2828.

\bibitem{Datta}
S. Datta, S. Gupta, M. Padmanath, J. Maiti and N. Mathur,
{\emph{JHEP} \bf{1302 (2013) 145}}

\bibitem{Pushkina}
I. Pushkina et al. [QCD-TARO Collaboration], {\emph{Phys. Lett. B} \bf{609 (2005) 265}},

\bibitem{Aarts:2015mma}
G.~Aarts, C.~Allton, S.~Hands, B.~J\"ager, C.~Praki and J.-I. Skullerud,
\textbf{Phys. Rev.} D92 (2015) 014503 [1502.03603]

\bibitem{Aarts:2015xua}
G.~Aarts, C.~Allton, S.~Hands, B.~J\"ager, C.~Praki and J.-I. Skullerud,
\textbf{PoS} LATTICE2015 (2015) 183 [1510.04040]

\bibitem{DeBoni:lat2016}
G.~Aarts, C.~Allton, D.~De~Boni, S.~Hands, B.~J\"ager, C.~Praki and J.-I. Skullerud,
\textbf{PoS} LATTICE2016 (2016) 037 [1610.07439]


\bibitem{Montvay:1994cy}
I.~Montvay and G.~M\"unster, \textit{Quantum Fields on a Lattice}
(Cambridge University Press, 1997) 1--491

\bibitem{Gattringer:2010zz}
C.~Gattringer and C.~B. Lang, \textit{Quantum Chromodynamics on the Lattice} (Springer, Heidelberg, 2010) 1--343

\bibitem{Leinweber:2004it}
D.~B. Leinweber, W.~Melnitchouk, D.~G. Richards, A.~G. Williams and J.~M.
  Zanotti,
\textbf{Lect. Notes Phys.} 663 (2005) 71--112

\bibitem{Edwards:2004sx}
{\scshape SciDAC, LHPC, UKQCD} collaboration, R.~G. Edwards and B.~Joo,
\textbf{Nucl. Phys. Proc. Suppl.} 140 (2005) 832

\bibitem{wip}
G.~Aarts, C.~Allton, D.~De Boni, S.~Hands, B.~J\"ager, C.~Praki and J.-I. Skullerud, \textit{Work in progress}

\bibitem{Asakawa:2000tr}
M.~Asakawa, T.~Hatsuda and Y.~Nakahara,
\textbf{Prog. Part. Nucl. Phys.}  46 (2001) 459--508






\bibitem{Aarts:2014cda}
G.~Aarts, C.~Allton, T.~Harris, S.~Kim, M.~P. Lombardo, S.~M. Ryan et~al.,
\textbf{JHEP}   07 (2014) 097

\bibitem{Aarts:2014nba}
G.~Aarts, C.~Allton, A.~Amato, P.~Giudice, S.~Hands and J.-I. Skullerud,
\textbf{JHEP}  02 (2015) 186

\bibitem{Amato:2013naa}
A.~Amato, G.~Aarts, C.~Allton, P.~Giudice, S.~Hands and J.-I. Skullerud,
\textbf{Phys. Rev. Lett.}  111 (2013) 172001

\bibitem{McClendon}
C. McClendon, \textit{Optimized Lattice QCD Kernels for a Pentium 4 Cluster}
Jlab preprint, JLAB-THY-01-29

\bibitem{Edwards:2008ja}
R.~G. Edwards, B.~Joo and H.-W. Lin,
\textbf{Phys. Rev.} D78 (2008) 054501


\bibitem{Lin}
H.-W. Lin \textit{et al.},
\textbf{Phys. Rev.}  D79 (2009) 034502

\bibitem{Gusken:1989ad}
S.~G\"usken, U.~L\"ow, K.~H. M\"utter, R.~Sommer, A.~Patel and K.~Schilling,
\textbf{Phys. Lett.} B227 (1989) 266--269

\bibitem{Albanese:1987ds}
{\scshape APE} collaboration, M.~Albanese et~al.,
\textbf{Phys. Lett.}   B192 (1987) 163--169



\bibitem{Yao:2006px}
{\scshape Particle Data Group} collaboration, W.~M. Yao et~al.,
\textbf{J. Phys.} G33  (2006) 1--1232

\bibitem{Durr}
S. D\"urr et al.,
\textbf{Journal Science} 322, (2008) 1224-1227


\bibitem{Agashe:2014kda}
{\scshape Particle Data Group} collaboration, K.~A. Olive et~al.,
\textit{Review of Particle Physics}, \textbf{Chin. Phys.} C38 (2014) 090001

\end{thebibliography}
\end{document}